\documentclass[aps,prl,reprint,floatfix,showpacs,groupedaddress,superscriptaddress]{revtex4-1}

\usepackage{bm}
\usepackage{xr}
\usepackage{epsfig}
\usepackage{comment}
\usepackage{epstopdf}
\usepackage{graphicx}
\usepackage{graphics}
\DeclareGraphicsExtensions{.eps}
\usepackage{color}

\usepackage{enumitem}

\usepackage{amsmath}
\usepackage{amsfonts}
\usepackage{amssymb}

\begin{document}

\title{Accurate Coulomb Blockade Thermometry up to 60 Kelvin}
\author{M.~Meschke}
\affiliation{Low Temperature Laboratory, Department of Applied Physics,
Aalto University School of Science, P.O. Box 13500, 00076 Aalto, Finland}
\author{A.~Kemppinen}
\affiliation{VTT Technical Research Centre of Finland Ltd, Centre for Metrology MIKES, P.O. Box 1000, FI-02044 VTT, Finland}
\author{J.P.~Pekola}
\affiliation{Low Temperature Laboratory, Department of Applied Physics,
Aalto University School of Science, P.O. Box 13500, 00076 Aalto, Finland}




\begin{abstract}
We demonstrate experimentally a precise realization of Coulomb Blockade Thermometry (CBT) working at temperatures up to 60 K. Advances in nano fabrication methods using electron beam lithography allow us to fabricate a uniform arrays of sufficiently small tunnel junctions to guarantee an overall temperature reading uncertainty of about 1\%
\end{abstract}


\maketitle

\section{Introduction}
Primary thermometers are required for the ongoing effort to redefine the Kelvin scale based directly on a fundamental physical constant, namely a fixed value of the Boltzmann constant $k_{\rm{B}}$ \cite{Fellmuth:2006}. 
This task requires primary thermometer realizations using different approaches, where consistency between different methods is compulsory. Noise thermometers, like the Magnetic Field Fluctuation Thermometer (MFFT)\cite{Enss:2014,Engert:2014}, Current Sensing Noise Thermometry (CSNT) \cite{Casey:2014} and the CBT \cite{Pekola:1994} are actually candidates that are under consideration for the low temperature range. Experimentally, the method of CBT is a well established tool for precise determination of temperatures below 1 K \cite{Pekola:1994, Anna:2012} with good precision \cite{Meschke:2012}. At higher temperatures though, fabrication inhomogeneities \cite{Hirvi:1995} are setting strict limits to the feasibility of CBT, so that no precise sensors operating above about 10 K were reported so far. We will present in this work a CBT operating at temperatures up to 60 K and discuss the uncertainty component arising from fabrication imperfections.  

\section{Theoretical background}

Coulomb Blockade Thermometry employs the temperature dependent electrical conductance $G$ of tunnel junction arrays. Classically, CBT works in a weak Coulomb blockade regime $E_{\rm{C}} \ll k_{B}T$, where the charging energy of the system with $N$ junctions in series is defined as $E_{\rm{C}}\equiv[(N-1)/N]e^2/C$, where $C$ is the capacitance of one island \cite{Pekola:1994}. The latter depends on the physical size of the tunnel junction in combination with the thickness of the insulator, in addition to the stray capacitance of the island. Note that temperature sensors for an increased temperature range need to have an increased charging energy and therefore smaller junctions. Temperature is derived in the strict limit of $E_{\rm{C}} \ll k_{B}T$ from the measurement of the full width of the conductance dip at half minimum using
\begin{equation} V_{1/2} \cong 5.439Nk_{B}T/e. 
\label{V12} 
\end{equation}
and for the limit $E_{\rm{C}} \simeq k_{B}T$ \cite{Farhangfar:1997,Anna:2012} with 
\begin{equation} 
V_{1/2}\cong5.439Nk_{B}T(1+0.3921\frac{\Delta G}{G_{T}})/e, 
\label{V12a} 
\end{equation}

In addition, a secondary temperature measurement can be obtained by recording the depth of the zero bias conductance dip 
\begin{equation} 
\frac{\Delta G}{G_{T}}=\frac{1}{6}u_{N}, 
\label{GGT} 
\end{equation} 
with the defined parameter $u_{N} \equiv E_{\rm{C}}/k_{B}T$.
Equation (\ref{GGT}) has to be extended by higher order contributions when leaving the strict limit of $E_{c} \ll k_{B}T$. The measured conductance follows well the expression\cite{Anna:2012}
\begin{equation} 
\frac{\Delta G}{G_{T}}=\frac{1}{6}u_{\rm{N}}-\frac{1}{60}u_{\rm{N}}^2+\frac{1}{630}u_{\rm{N}}^3. 
\label{GGT3} 
\end{equation} 

We use in this work a numerical \cite{Pekola:1994} way to precisely calculate the conductance curves that includes no approximation at any ratio of $E_{c}$ to $ k_{B}T$ in order to extend the description into the intermediate Coulomb blockade regime ($E_{\rm{C}}\simeq k_{B}T$) \cite{Anna:2012}. This allows us to calculate conductance curves also including the overheating effects \cite{Meschke:2004}.

\section{Fabrication of the suspended germanium mask}

We fabricate the devices using electron beam lithography combined with shadow evaporation technique \cite{Niemeyer:1976,Dolan:1977} and a special tri-layer resist scheme \cite{Pashkin:2000,Pekola:2013} that employs in our case a 22 nm thin germanium (Ge) film as the material for the suspended mask. The wafer preparation is done as follows: at first, we spin coat (5500 rpm for 60 s) a 4 inch silicon wafer covered with 300 nm of silicon oxide with a 400 nm thick copolymer film (11 \% MMA/PMMA in ethyl lactate) and bake it for 45 min at 175 degree Celsius. This extremely long and hotter than usual baking time is required to avoid out-gassing of the copolymer underneath the gas-tight Ge layer that is deposited next. We use an electron gun evaporator equipped with a powerful cryo--pump to deposit the 22 nm thin Ge film with a growth rate of 0.1 \AA/s in order to achieve a low stress Ge film. This corresponds to a total evaporation time of about 22 minutes for the 22 nm film. We have to control the deposition rate precisely since it has a strong effect on the strain of the Ge film, which can vary from positive to negative. The optimal rate is stable in our evaporator, but in general depends on the vacuum conditions. We observe that more perfect vacuum conditions require a lower deposition rate. Finally, we again spin coat the wafer with a 50 to 60 nm thick standard PMMA electron beam resist (molecular weight 2200k) using a 2.25 \% solution in anisole at 2500 rpm and bake at 150 degree Celsius for 90 seconds. This procedure yields a very reproducible quality of the resist stack and allows us to achieve a constant resolution and device quality over the full 4 inch wafer surface. We perform the electron beam lithography with an e-beam writer having 100 keV electron energy, doses of 1000 $\mu$C/cm$^2$, no proximity correction and beam currents of typically 1 nA. 

The development for the mask is separated into three phases\cite{Pashkin:2000,Pekola:2013}: firstly, we develop the exposed top layer of PMMA using standard MIBK:IPA 1:3 solution, followed by pure IPA bath and finally rinsing in deionized water, 30 seconds for each process step. Secondly, we transfer the opening in the PMMA to the germanium film using reactive ion etching (RIE) with $\rm{CF}_4$ plasma at a pressure of 40 mTorr and a rf-power of 40 W for 140 s. The $\rm{CF}_4$ plasma selectively etches only the germanium film and does not affect the resist noticeably. Thirdly, we use pure oxygen plasma at a pressure of 30 mTorr and a rf-power of 40 W for 15 min that an--isotropically etches selectively the copolymer film and removes at the same time the PMMA film covering the remaining germanium mask. Increasing the pressure of the oxygen plasma to 225 mTorr for additional 20 min etches more isotropically and creates the necessary undercut below the germanium mask, enabling shadow evaporation at an angle of about $\pm$ 25$^\circ$. 

\begin{figure}[!h]
\centering\includegraphics[width=0.48\textwidth]{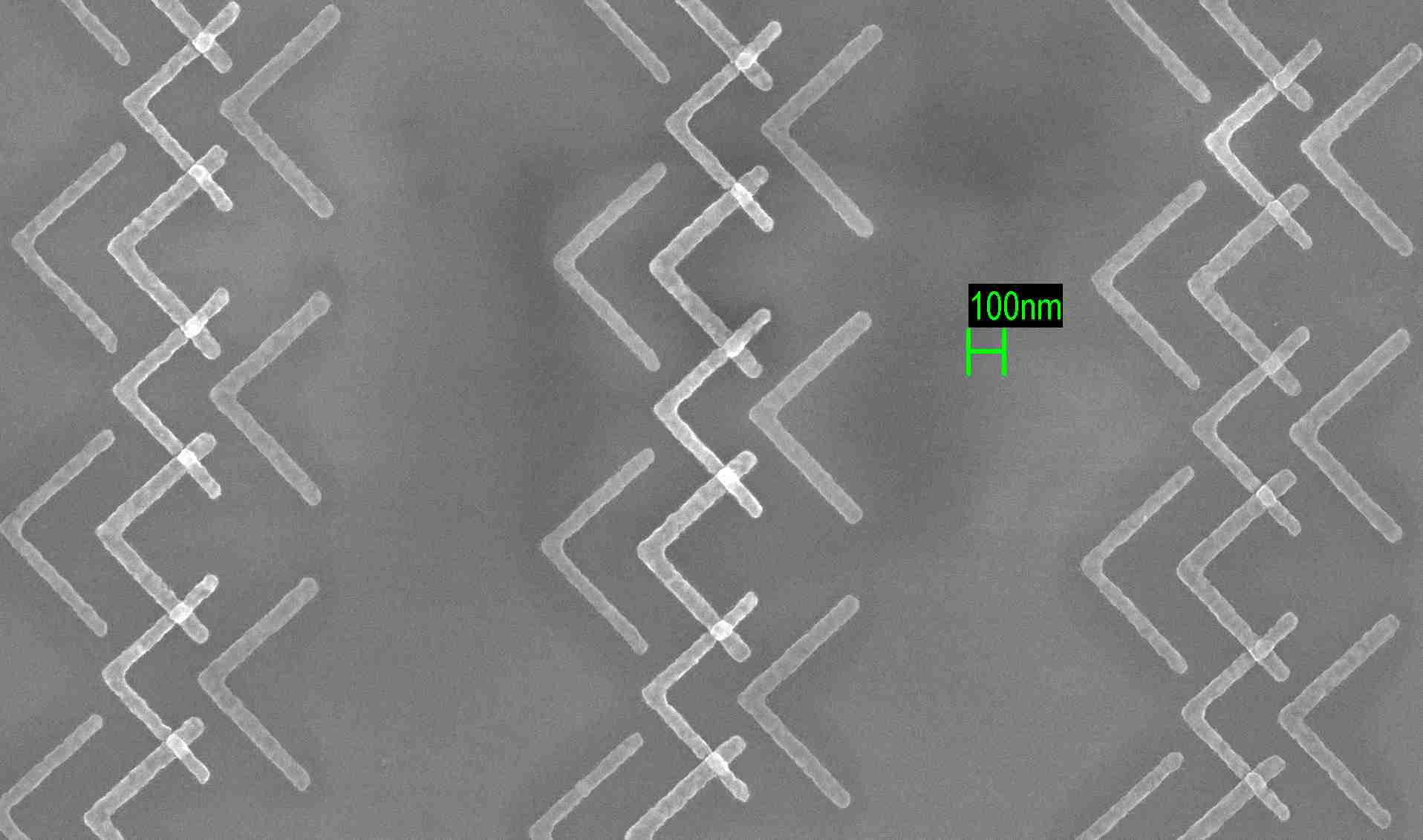}

\caption{A scanning electron micrograph of a sample identical to the measured one. Depicted are three rows of 6 junctions each fabricated at two angle shadow evaporation (see text). We observe a remarkable reproducibility over the full 77x30 junction array, note that the 2nd shadow, that falls in the image always on the left, has a slightly reduced size due to partial blocking of mask from the first angle evaporation.)}
\label{real_sensor}
\end{figure}

The next step in the fabrication is the shadow evaporation performed in an electron gun evaporator: we deposit first under a tilt angle of + 20$^\circ$ a 20 nm thin aluminium film with a high deposition rate of about 12 \AA/s. The high deposition rate produces a film with a reduced grain size. Then, we oxidise the aluminium film with 5 mbar of pure oxygen in the deposition chamber for 300 s to create the AlOx tunnel barriers. Finally, we deposit the second aluminium film with a tilt angle of -20$^\circ$. The lift--off is done by immersing the wafer in acetone at ambient temperature. 
Figure \ref{real_sensor} depicts a small section of one sensor that consists in total of an array of 77 junctions in series to increase the signal strength and to suppress the influence of the electromagnetic environment \cite{Hirvi:1995,Pekola:2008} and in 30 parallel rows to reduce the final resistance of the sensor to a value of about 100 k$\Omega$ in order to facilitate the electrical measurement.  
Already the design of the mask takes the main requirement of the sensor, the small junction size in combination with a small scatter in the junction areas into account: the crossing geometry, where the second shadow overlaps the first one totally defines the junction area leaving only the line width fluctuations as an error source. Even smaller junction areas are possible to fabricate if the second shadow would only partly cover the first one, but in that case, both the length of the line and a misalignment of the mask with respect to the tilting angle would contribute to the error in defining the junction size and consequently the important tunnel junction resistance. 
As depicted in Fig. \ref{real_sensor}, the final junction size of about 45 nm x 32 nm is constant over all sensor junctions, the material deposited for the first layer blocks partly the mask and yields the reduced line--width of the second shadow.

\section{Nonlinear background correction for the operation of CBT sensors at elevated temperatures}

\begin{figure}[!h]
\centering\includegraphics[width=0.48\textwidth]{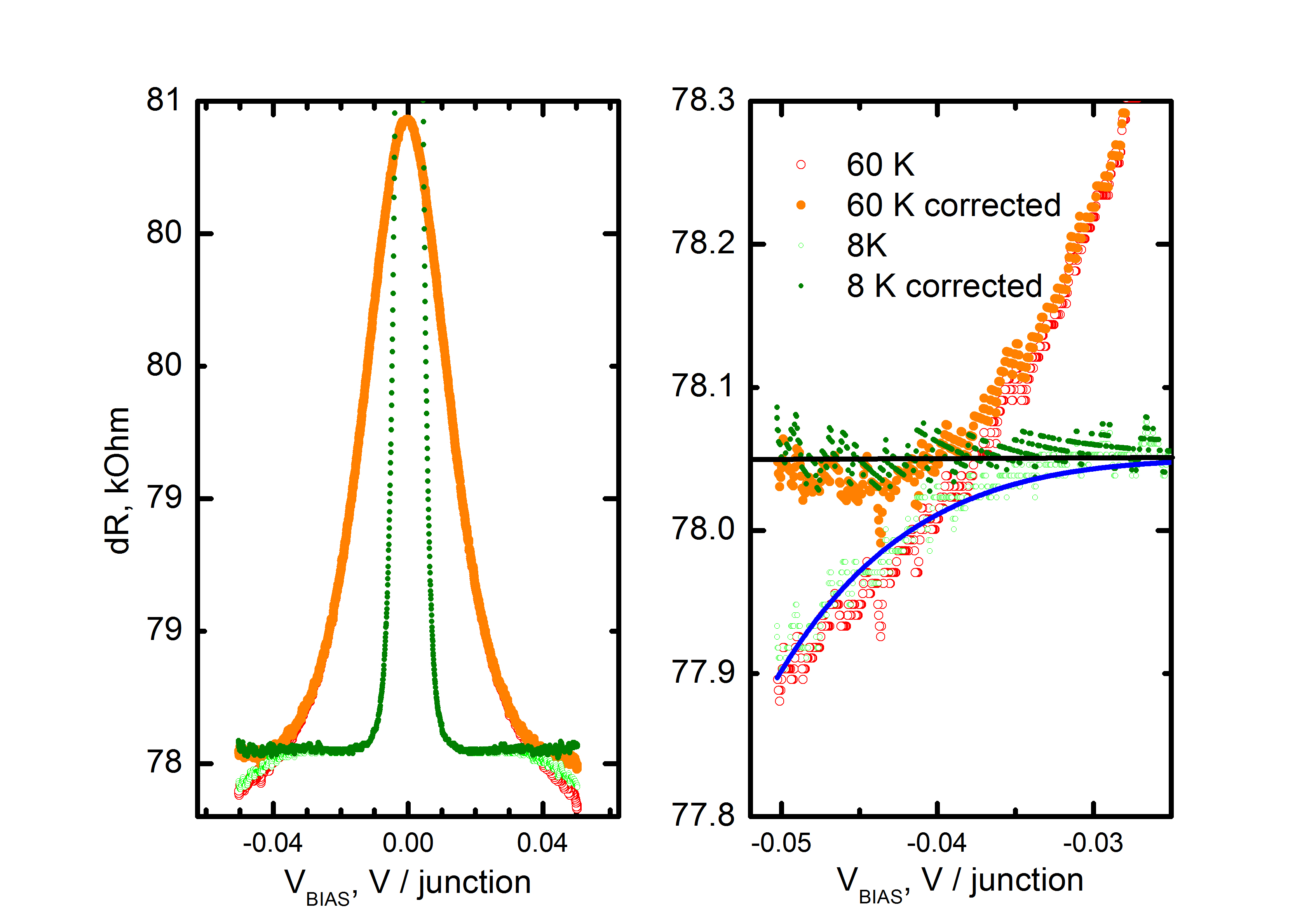}
\caption{(left) Measured voltage dependent differential resistance curves at two different temperatures for a CBT sensor. The curve measured at 8 K was clipped to emphasize the non--linear background, the zero bias value reaches 95 k$\Omega$. Both uncorrected and corrected curves are depicted. (right) Zoom into the area affected by non--linear background, the blue line is the function describing the background and the black horizontal line is a guide to the eye at the position of the tunnel junction resistance for small bias values.}
\label{background}
\end{figure}

The measured total sensor resistance is 78 k$\Omega$ (see Fig. \ref{background}) corresponding to a resistance per junction of 200 k$\Omega$. A practical realization of a CBT sensor shall show a linear voltage dependence up to applied bias voltages that are ideally few times the half width of the Coulomb peak. At low enough temperatures, the assumption of a bias voltage independent conductance (apart from the Coulomb blockade peak) of the aluminium oxide tunnel barriers holds well enough: the Coulomb peak is well separated from the non--linear background even at a temperature of 8 K (see Fig. \ref{background}). 

Real devices show due to the finite tunnel barrier height (about 2 eV \cite{Gloos:2000} in the case of a tunnel barrier made from AlOx) noticeable deviations at higher voltages: such a finite barrier height leads according to the phenomenological Simmons model \cite{Simmons:1964} to a parabolic deviation in the form $G(V)=G_{\rm{0}}(1+(V/V_{\rm{0}})^2)$ with a typical value of $V_{\rm{0}}=0.25 V$\cite{Gloos:2000,Gloos:2003}. Aiming at a temperature reading of 60 K requires a bias voltage range of $\pm V_{\rm{BIAS}}\approx 2V_{\rm{1/2}}\approx 50 $ mV per junction as increasing the temperature also increases linearly $V_{\rm{1/2}}$, where the simple model predicts a noticeable deviation on the order of 1\%, whereas the peak-hight itself is reduced to about 3\% in the sensors discussed here. Experimental findings differ somewhat from the Simmons model, even a more sophisticated model\cite{Jung:2009} employing first principle calculations to describe the density of states in the tunnel barrier catch well the background at high voltage, but falls short in describing the detailed behaviour close to the region of interest here. As a solution, we determine the background using the depicted curve measured at 8 K (see Fig. \ref{background}) by a polynomial function in the form of $R=R_0(1+\alpha V^6)$ (see Fig. \ref{background} (right)) and correct the measured curves with this function at all temperatures. An alternative way to limit the fitting to the range where the background is flat (+/- 30 mV/junction) reveals within the confidence interval of the fitting routine no deviation between the two methods. Figure \ref{background} (right) shows, that the course of the background is very similar for different temperatures and the applied correction brings both curves to a constant value within the noise level of the measurement. In conclusion, we show in this section that we can neglect the uncertainty component arising from the background at low enough temperatures ($\simeq$ 40 K), as the background does not deviate noticably in this range. Furthermore, the presented background correction is effective at temperatures between 40 K and 60 K.

\section{Experimental characterization of the multi tunnel junction sensors}

We use a standard setup\cite{Meschke:2012} including lock--in amplifiers to measure the bias dependent differential conductance curves of the devices. The sensors are wire bonded to a sample holder that is mounted in the vacuum chamber of a dilution refrigerator immersed in a liquid helium dewar. The dilution refrigerator is operated without condensed $^3$He/$^4$He mixture in order to access the temperature range up to 60 K: we apply heat to the sample stage with the standard electrical heater that overcomes the cooling towards 4 K due to the residual heat exchange gas in the vacuum chamber and the thermal conduction of the apparatus.

\begin{figure}[!h]
\centering\includegraphics[width=0.48\textwidth]{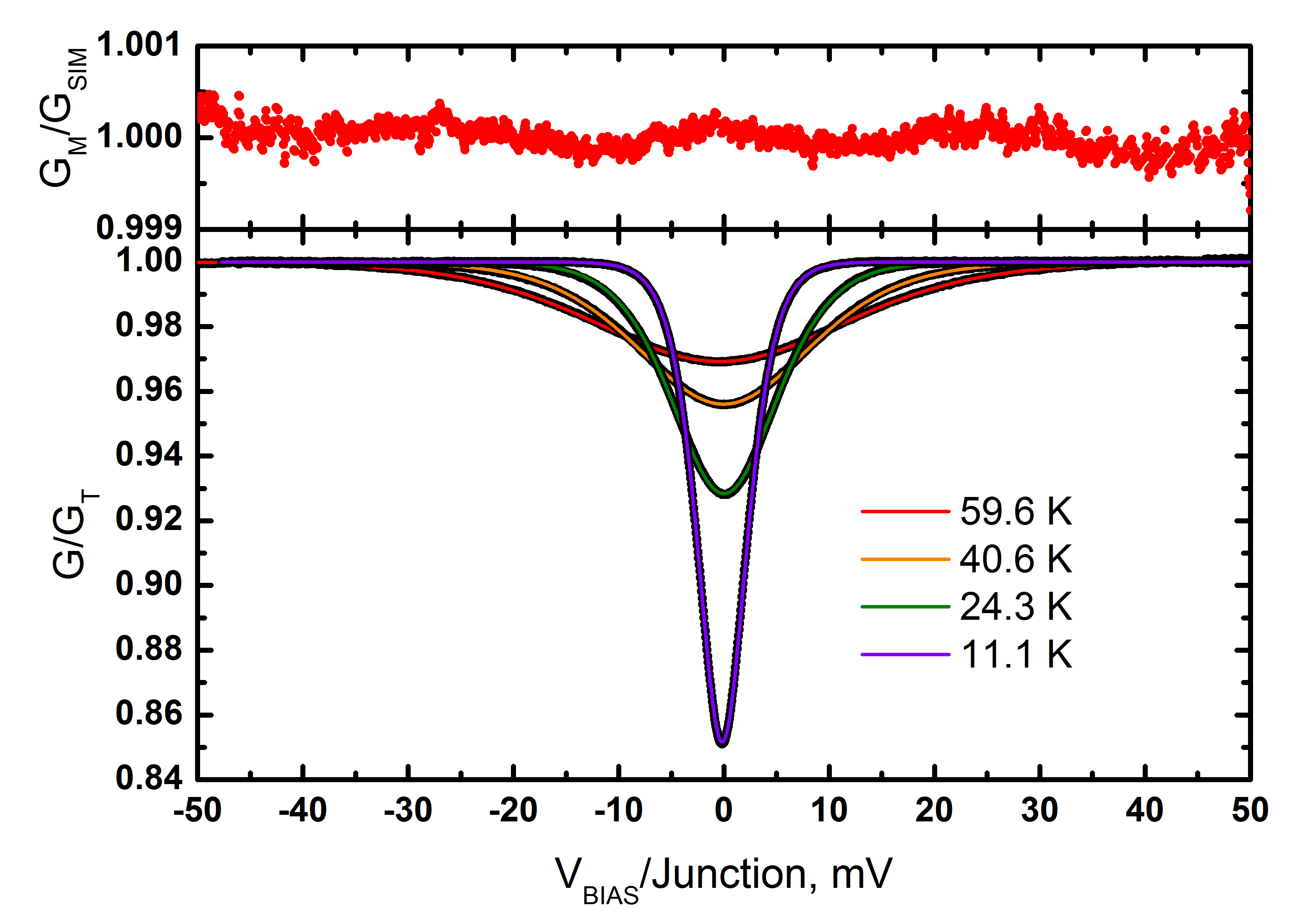}

\caption{Measured (black dots) and numerical fits (lines, see text) of bias voltage dependent, normalized conduction values of a CBT sensor at four different temperature values between 10 and 60 K. The voltage axis is scaled to depict the average voltage drop across one single junction in the sensor array. The top panel depicts the ratio between the measured ($G_{\rm{M}}$) and the simulated ($G_{\rm{SIM}}$) conductance at 59.6 K.}
\label{fig_CBT_ggt}
\end{figure}

Fig. \ref{fig_CBT_ggt} depicts few measured conduction curves together with the extracted temperature values from the fit. Despite the unusually high temperatures, the model still has to take overheating effects into account \cite{Meschke:2004} as the volume of the metallic islands is quite small (about 5e-4 $\mu$ m$^3$) and the applied voltage and consequently joule heating is increased: At a base temperature of 9 K, the electronic temperature in the islands rises at the applied voltage of $V_{\rm{1/2}}$ by about 100 mK. Already the perfect agreement between the theoretical model and the measured data, the magnitude of the signal even at exceptionally high temperature of 60 K and the flat background underline the outstanding quality of these devices. We find a charging energy as high as $E_{\rm{C}}/k_{\rm{B}}=10.6$ K, meaning that the sensors operate even at this high temperatures mainly in the intermediate Coulomb blockade regime\cite{Anna:2012}. This is very beneficial for the signal magnitude and for suppressing the remaining influence of the non--linear background.

\section{Experimental characterization of the sensor uniformity}

The main topic of this chapter is the experimental characterization of the fabricated multi tunnel junction sensors with respect to the resistance uniformity between the individual junctions. The latter is too difficult to probe directly because of the mere number of junctions and their small size. Instead, we follow here four approaches to determine this value experimentally and show that all yield a consistent result: (a) we fabricate together with the real device a test structure that enables to probe 11 individual test junctions, (b) measure three sensors on the very same chip in a range of temperatures to analyze the deviations, (c) compare the sensor to a reference temperature when immersed in liquid helium and finally (d) compare the temperature reading of the device to a single tunnel junction thermometer.

\subsection{Probing of individual junctions}
\label{ssec:best}

\begin{figure}[!h]
\centering\includegraphics[width=0.48\textwidth]{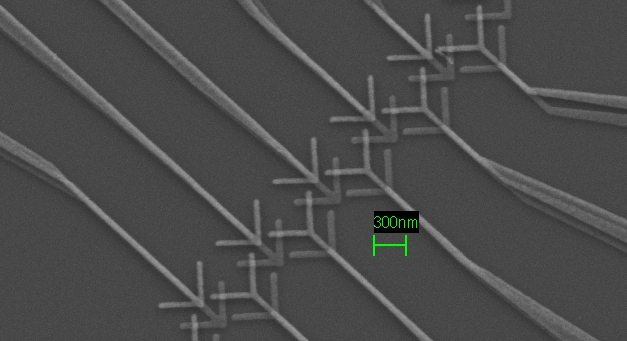}
\caption{Scanning electron mikrograph of an array of 11 junctions connected to 12 leads allowing electrical probing of all junctions individually. The junction size and layout is identical to the size of the structures in the sensors.}
\label{probe}
\end{figure}

We performed a direct measurement of the tunnel junction resistance at room temperature using the structure depicted in Fig. \ref{probe}, that is fabricated along with the real structures and allows monitoring of the fabrication accuracy. The 12 leads connect to bonding pads that are wire bonded to the test set--up.  

The immediate environment close to the junctions on the chip during fabrication of the test junction array differs somewhat from the real sensor (compare to Fig. \ref{real_sensor}) due to the additional leads, but the inspection with a scanning electron microscope reveals no significant influence on the junction area or quality of the mask. Consequently, we assume that we can use the results of the test structures to model the full sensor array: We find in the best case a mean resistance of 70 k$\Omega$ with a standard deviation of 3.3 k$\Omega$, corresponding to 4.7\%. We use this result to simulate the deviation of a full sensor consisting of 77 junctions using the method described in \cite{Hirvi:1995,Hahtela:2013} and find a resulting temperature deviation of 0.07\%. As an upper limit, we found a resistance deviation of 11.4\%, still resulting in a temperature deviation of 0.7\%, well below 1\%. 
These findings allow under the given assumptions to set an upper limit of the uncertainty component of temperature determination arising from fabrication non--uniformities to 1\%.

\subsection{Inter--comparison of three sensors}

We describe in what follows the inter--comparison of three CBT sensors in the temperature range between 10 K and 60 K. The result is at the first glance already very promising: the three sensors agree with each other on a level of about 0.2\% over the investigated temperature range. 

\begin{figure}[!h]
\centering\includegraphics[width=0.48\textwidth]{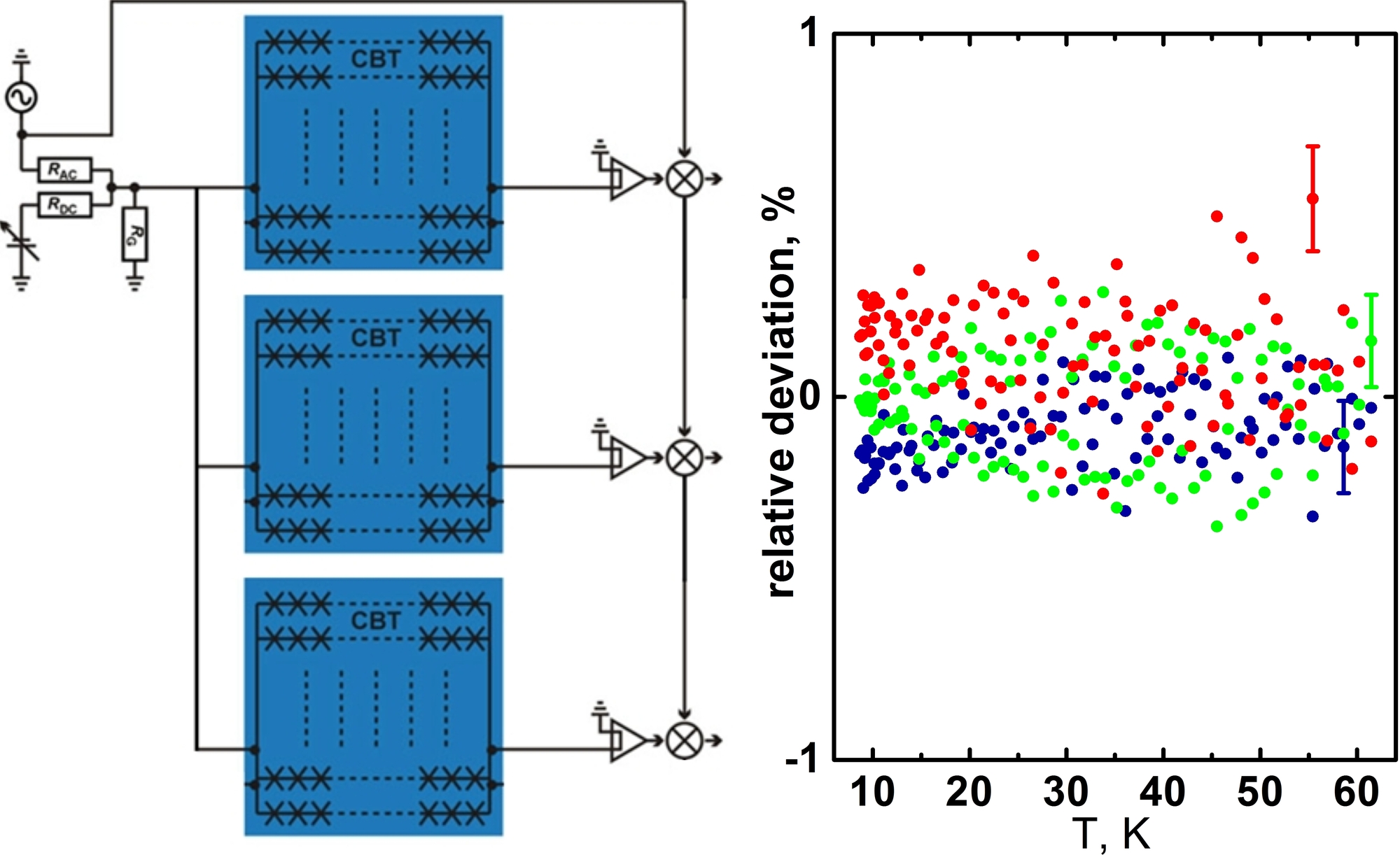}
\caption{(left) Experimental set--up measuring three CBT sensors simultaneously: one single voltage source is attached to three sensors. The voltage consists of two components, the resistance network adds a divided DC bias voltage set by $R_{\rm{G}}/R_{\rm{DC}}$ and a sinusoidal AC component via $R_{\rm{G}}/R_{\rm{AC}}$. The resulting differential conductance of the sensor is measured using one current amplifier and one lock--in amplifier each. Only the sensors (blue shaded area) are located at low temperature and connected to room temperature via filtered Thermocoax lines. (right) Resulting relative deviation among three sensors (red, blue, green circles) with respect to the measured mean temperature value of three sensors in the temperature range between 10 and 60 K. Given exemplary error bars are the one sigma confidence interval of the fit to the measured conductance curves and agree well with the observed scatter of the temperature data.}
\label{3cbt}
\end{figure}

The setup  (Fig. \ref{3cbt} left) is specially optimized in order to rule out other contributions to the measurement error: the very same single voltage relating the measurement to temperature is applied to all three sensors that are measured at the same time. In addition, all three sensors are fabricated together with single junction devices on one single silicon wafer leaving only little room for temperature gradients between the sensors. Finally, as the setup allows to measure the three sensors at the same time, any temperature fluctuations or drifts affect all sensors identically. This leaves only the error source arising from the sensors itself influencing the results, if one assumes in addition no significant deviations from linearity in the amplifier chain.

We discuss two further questions at this point: (a) what  is the expected absolute uncertainty that is consistent with the observed relative deviation? And (b) to what upper limit of junction array inhomogeneity does this value correspond to? 
In order to answer those questions, we follow the approach described in \cite{Hahtela:2013}: we calculate numerically the conductance curves for arrays that have a random fluctuation of resistances in the individual junctions characterized by a Gaussian distribution with a set of different standard deviations. Now we can link the expected absolute deviation of temperature reading for a sensor to the relative deviation among many sensors with different individual random fluctuations of resistances. The result is that a 10\% scatter in the junction resistance (corresponding to about 0.7\% absolute deviation in temperature reading) would lead to a 0.2\% deviation among different sensors and describe best the experimental findings. These results are again consistent with an upper limit of the uncertainty of temperature determination arising from fabrication non-uniformities to be smaller than 1\%.

\subsection{Comparison of the sensor reading towards a reference temperature}

We demonstrate in this section a comparison of the sensor reading to the reference temperature realized with a liquid helium bath in combination with measuring the vapor pressure of the cryo--liquid: the sensor is immersed in a pressurized liquid helium dewar at 1150 mbar, corresponding to a temperature of 4.359 K according to the ITS-90 temperature scale. We neglect at this point temperature gradients within the liquid helium and the uncertainty arising from the pressure measurement ($\pm$ 10 mbar), as the latter corresponds to a temperature deviation of < 0.2 \%.
 
\begin{figure}[!h]
\centering\includegraphics[width=0.48\textwidth]{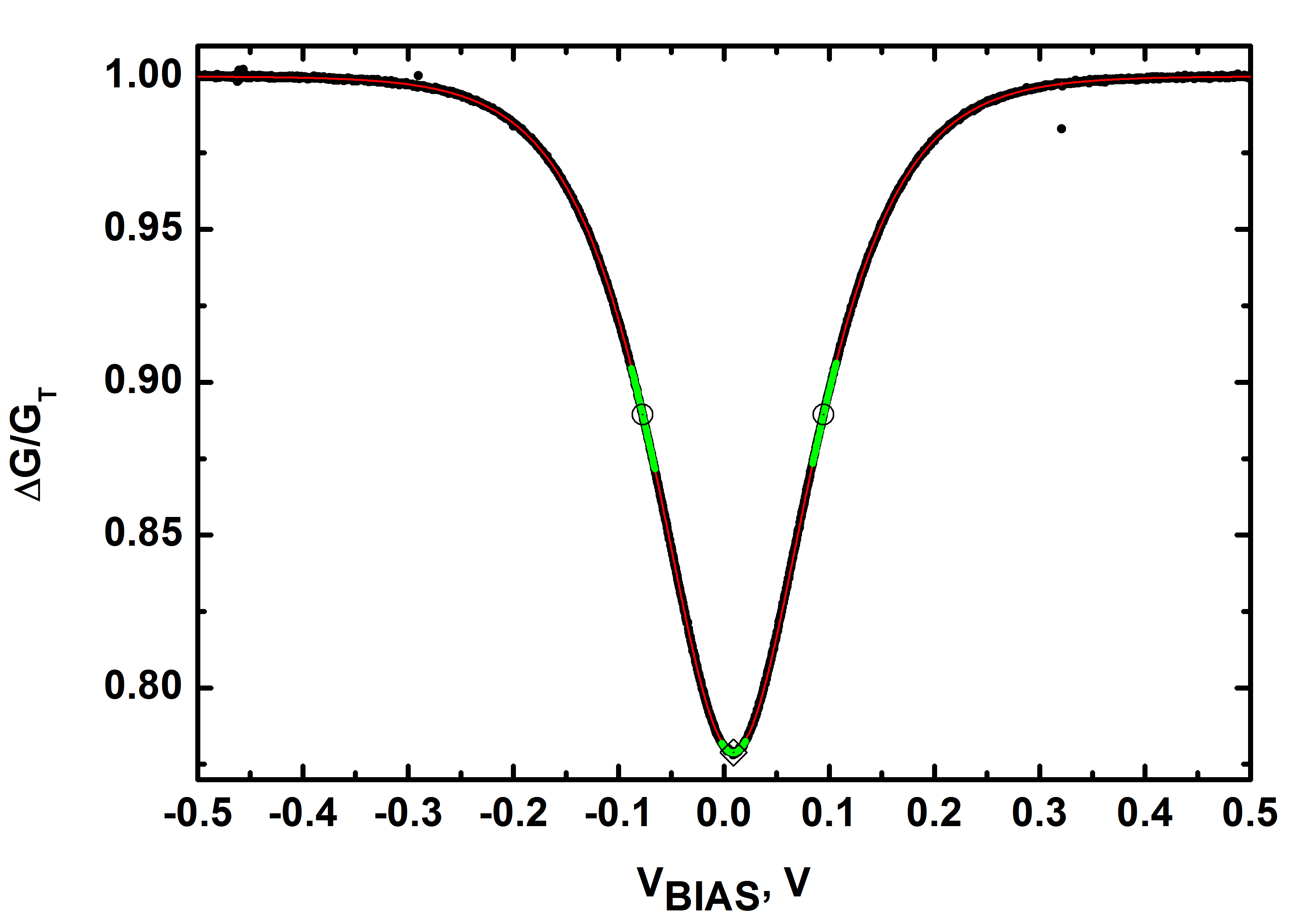}
\caption{Voltage bias dependent conductance curve (dots) of a sensor immersed in liquid helium with a fit to the full model (red line). The green lines mark the intervals, where a simple polynomial fit of second order is used to determine the half width of the dip. Open circles mark the position of the half value of the conductance dip on the curve, the rhombus marks the conductance minimum (see text).}
\label{Helium}
\end{figure}

Temperature is extracted from the conductance curve measurement in two ways: fitting the numerically calculated conductance curve to the full data range (4.363 K) and by determining temperature via the half width of the dip in the conductance curve using second order polynomial fits to the dip and around the two values neat half minimum as indicated in Fig. \ref{Helium}, to determine temperature using Eq. (\ref{V12a}) (4.385 K). Note, that the agreement between all values is within 0.5\% and that there is an excellent match between measured data and the theoretical curve. These results are again consistent with an upper limit of the uncertainty of temperature determination arising from fabrication inhomogeneities to be smaller than 1\% for the presented sensors. Note that the charging energy of the sensor presented in this chapter was lower (5.5 K) in order to adapt the conductance dip magnitude to the operation temperature close to 4.2 K. Sensors having the higher charging energy discussed above would already suffer from the influence of random background charge distributions at this temperature \cite{Anna:2012}.

\subsection{Experimental comparison to a single junction thermometer}

Finally, we discuss the comparison of the CBT sensor with a single junction device (SJT, \cite{Pekola:2008,Meschke:2012,Chen:2014}). The obvious advantage of the SJT over a multi junction device is that no scatter of junction resistance in the array influences the result. Like this, the direct comparison of the CBT and SJT sensor shall reveal the influence of the non--uniformity. We choose an optimal temperature of about 9 K for the experiment, where on the one hand the overheating effects and on the other hand the influence of the non--linear background are still negligible.

\begin{figure}[!h]
\centering\includegraphics[width=0.48\textwidth]{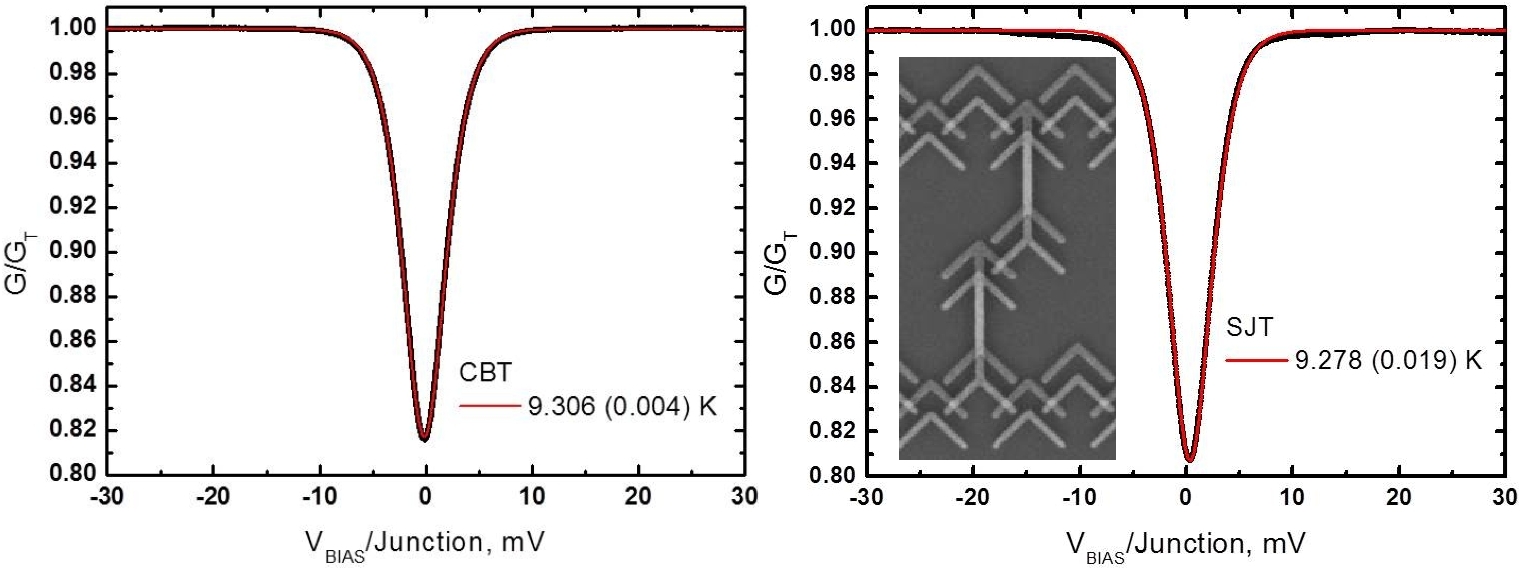}
\caption{Measured (black dots) and fitted (red line) bias voltage dependent conductance curves for a CBT (left) and a single junction (right) at a temperature of about 9.3 K. The uncertainties of the temperature readings given in brackets are the one sigma confidence interval of the fit to the measured conductance curves and neglect other components (see text). The voltage axis of the CBT measurement is scaled to that of a single junction. The inset of the SJT graph shows a scanning electron micrograph of a device that is similar to the measured one (see text for details).}
\label{cbt_sjt}
\end{figure}
An image of the SJT is depicted in Fig. \ref{cbt_sjt}, left inset: one central junction is connected via four long arrays of tunnel junctions allowing a four probe measurement of the central junction and creating a suitable electromagnetic environment for precise thermometry \cite{Pekola:2008}. We use a standard setup \cite{Meschke:2012} for the measurement of the differential conductance; the main difference with respect to the CBT setup is a voltage amplifier that is needed to detect the applied bias voltage across the single junction.  
The main disadvantage, the small signal magnitude, of using only one junction is partly compensated by an increased time constant of the lock--in amplifiers, leading to extremely long measurement times for a single data point of about 12 hours in order to yield comparable signal to noise ratios for both sensor types. The latter causes some long term drift effects in the background of the SJT conductance curve (Fig. \ref{cbt_sjt}, right), most likely due to amplifier gain deviations caused by room temperature changes. 
The agreement of the two temperature readings is perfect. One has to consider the main uncertainty components though: the voltage axis realization of the SJT relies on the accuracy of the amplifier gain that is limited to 1\% whereas the main influence in the case of CBT comes from the sensor non--uniformity. This comparison can set again an upper limit of the accuracy for the described CBT sensors to 1\%.

\section{Conclusion}

We demonstrate in this paper an experimental realization of CBT sensors with high charging energy suitable for relating temperature of up to 60 K to the Boltzmann constant with a precision that is better than one percent. The demonstrated fabrication scheme allows a sufficient fabrication accuracy to reduce the inhomogeneity in the parameters of the 
tunnel junction array in order to extend the useful temperature range of multi tunnel--junction devices beyond earlier realizations \cite{Tilke:2003} and expectations \cite{Chen:2014}. The results show that CB thermometry is indeed feasible at temperatures, where earlier only single junction devices were available\cite{Chen:2014,Pekola:2008}, enhancing now the performance of the sensors due to the superior signal to noise ratio, allowing for faster signal acquisition or higher precision. These advances have also consequences for sensors operating in the classical temperature range between 50 mK and 1 K: if one fabricates junctions with a larger area of about of 1x1 $\mu$m$^2$ suitable for this temperature range with the same process and absolute fabrication accuracy, the resulting variation of the total junction area will be negligible. Other effects, like variations in the oxide layer thickness\cite{Gloos:2003} would become dominating in this case and would need to be optimized.

We focus in this paper on two uncertainty components that are crucial for high-temperature measurements: the background correction, where we show that the uncertainty component can be made negligible due to an increased charging energy and the array uniformity where we demonstrate an upper limit of 1\% in this work. The whole uncertainty budget was not assessed. Other important uncertainty components are the fitting uncertainty due to measurement noise and disturbances and finally the measurement of the dc voltage in the G(V) measurements. The latter can be suppressed roughly to 0.01\% \cite{Hahtela:2015}. 
We can demonstrate in this work only an upper limit for the uncertainty, but our results would allow the conclusion that the presented sensors are even more accurate. A direct comparison to a realization of the ITS90 temperature scale in combination with a traceable voltage measurement scheme would allow a characterization with an one order of magnitude smaller uncertainty. 

CB Thermometry turns out to be a mature technology operating nowadays in a temperature range spanning over more than four decades in temperature from 60 K as shown in this work down to the low milli--kelvin range \cite{Scheller:2014}, reaching electronic temperatures as low as 3.7 mK if the sensor is immersed directly in liquid helium of a dilution refrigerator\cite{Prance:2015}.

\subsection{Acknowledgements}
{We would like to thank Yu. Pashkin and S. V. Lotkhov for valuable inputs. 
We acknowledge the availability of the facilities and technical
support by Otaniemi research infrastructure for Micro
and Nanotechnologies (OtaNano).
We acknowledge
financial support from the EMRP
(project no. SIB01-REG2) and the Academy of Finland
though its LTQ CoE grant (project no. 250280) and the project no. 259030 of the Academy of Finland.}


\begin{thebibliography}{9}




\bibitem{Fellmuth:2006} Fellmuth, B., Gaiser, C. \& Fischer, J. 2006 Determination of the Boltzmann constant--status and prospects. \textit{Meas. Sci. Technol.} \textbf{17}, R145-R159.

\bibitem{Enss:2014} Rothfu{\ss}, D., Reiser, A. Fleischmann, A. \& Enss, C. 2014 A Microkelvin Magnetic Flux Noise Thermometer. \textit{J. Low Temp. Phys.} \textbf{175}, 776-783.

\bibitem{Engert:2014} Kirste, A., Regin, M, Engert, J, Drung, D. \& Schurig, T. 2014 A calculable and correlation-based magnetic field fluctuation thermometer. \textit{Journal of Physics: Conference Series} \textbf{568}, 032012.

\bibitem{Casey:2014} Casey, A., Arnold, F., Levitin, LV., Lusher, CP., Ny\'eki, J., Saunders, J., Shibahara, A., van der Vliet, H., Yager, B., Drung, D., Schurig, T., Batey, G., Cuthbert, MN. \& Matthews, AJ. 2014 Current Sensing Noise Thermometry: A Fast Practical Solution to Low Temperature Measurement. \textit{J. Low Temp. Phys.} \textbf{175}, 764-775.

\bibitem{Pekola:1994} Pekola, JP., Hirvi, KP., Kauppinen, JP. \& Paalanen, MA. 1994 Thermometry by arrays of tunnel junctions. \textit{Phys. Rev. Lett.} \textbf{73}, 2903-2906.

\bibitem{Anna:2012} Feshchenko, AV., Meschke, M., Gunnarsson, D., Prunnila, M., Roschier, L., Penttil\"{a}, JS. \& Pekola, JP. 2012 Primary Thermometry in the Intermediate Coulomb Blockade Regime. \textit{J. Low Temp. Phys.} \textbf{173}, 36-44.

\bibitem{Meschke:2012} Meschke, M., Engert, J., Heyer, D. \& Pekola, JP. 2011 Comparison of Coulomb Blockade Thermometers with the International Temperature Scale PLTS-2000. \textit{Int. J. Thermophys.} \textbf{32}, 1378-1386.

\bibitem{Hirvi:1995} Hirvi, KP., Kauppinen, JP., Korotkov, AN., Paalanen, MA. \& Pekola, JP. 1995 Arrays of normal metal tunnel junctions in weak Coulomb blockade regime. \textit{Appl. Phys. Lett.} \textbf{67}, 2096-2098.

\bibitem{Farhangfar:1997} Farhangfar, S., Hirvi, KP., Kauppinen, JP., Pekola, JP., Toppari, JJ., Averin, DV. \& Korotkov, AN. 1997 One dimensional arrays and solitary tunnel junctions in the weak coulomb blockade regime: CBT thermometry. \textit{J. Low Temp. Phys.} \textbf{108}, 191-215.

\bibitem{Meschke:2004} Meschke, M., Pekola, JP., Gay, F., Rapp, RE. \& Godfrin, H. 2004 Electron thermalization in metallic islands probed by coulomb blockade thermometry. \textit{J. Low Temp. Phys.} \textbf{134}, 1119-1143.

\bibitem{Niemeyer:1976} Niemeyer J. \& Kose V. 1976 Observation of large dc supercurrents at nonzero voltages in Josephson tunnel junctions. \textit{Appl. Phys. Lett.} \textbf{29}, 380-382.

\bibitem{Dolan:1977} Dolan, GJ. 1977 Offset masks for lift-off photoprocessing. \textit{Appl. Phys. Lett.} \textbf{31}, 337-339.

\bibitem{Pekola:2013} Pekola, JP., Saira, OP., Maisi, VF., Kemppinen, A., M\"{o}tt\"{o}nen, M., Pashkin, YA. \& Averin, DV. 2013 Single-electron current sources: Toward a refined definition of the ampere. \textit{Rev. Mod. Phys.} \textbf{85}, 1421-1472.

\bibitem{Pashkin:2000} Pashkin, YA., Nakamura, Y. \& Tsai, JS. 2000 Room--temperature Al single--electron transistor made
by electron--beam lithography. \textit{Appl. Phys. Lett.} \textbf{76}, 2256-2258.

\bibitem{Pekola:2008} Pekola, JP., Holmqvist, T. \& Meschke, M. 2008 Primary Tunnel Junction Thermometry. \textit{Phys. Rev. Lett.} \textbf{101}, 206801.

\bibitem{Gloos:2000} Gloos, K., Poikolainen, RS. \& Pekola, JP. 2000 Wide-range thermometer based on the temperature-dependent conductance of planar tunnel junctions. \textit{Appl. Phys. Lett.} \textbf{77}, 2915-2917.

\bibitem{Simmons:1964} Simmons JG. 1964  Generalized Thermal J--V Characteristic for the Electric Tunnel Effect. \textit{J. Appl. Phys.} \textbf{35}, 2655-2658.

\bibitem{Gloos:2003} Gloos, K., Koppinen, PJ. \& Pekola, JP. 2003. Properties of native ultrathin aluminium oxide tunnel barriers. \textit{ J. Phys. Cond. Matt.} \textbf{15}, 1733-1746.

\bibitem{Jung:2009} Jung, H., Kim, Y., Jung, K., Im, H., Pashkin, YuA., Astafiev, O, Nakamura, Y, Lee, H., Miyamoto, Y. \& Tsai, JS. 2009 Potential barrier modification and interface states formation in metal-oxide-metal tunnel junctions. \textit{Phys. Rev. B} \textbf{80}, 125413.

\bibitem{Hahtela:2013} Hahtela, OM, Meschke, M, Savin, A, Gunnarsson, D, Prunnila, M, Penttil\"{a}, JS, Roschier, L, Heinonen, M, Manninen, A. \& Pekola, JP. 1995 Investigation of uncertainty components in Coulomb blockade thermometry. \textit{AIP Conf. Proc.} \textbf{1552}, 142-147.

\bibitem{Chen:2014} Chen, IH., Lai, WT. \& Lia, PW. 2014 Realization of solid--state nanothermometer using Ge quantum--dot
single--hole transistor in few--hole regime. \textit{Appl. Phys. Lett.} \textbf{104}, 243506.


\bibitem{Tilke:2003} Tilke, AT., Pescini, L, Lorenz, H \& Blick, RH. 2003 Fabrication and transport characterization of a primary thermometer formed by Coulomb islands in a suspended silicon nanowire. \textit{Appl. Phys. Lett.} \textbf{82}, 3773-3775.

\bibitem{Hahtela:2015} Hahtela, O. et al. (in preparation).

\bibitem{Scheller:2014} Scheller, CP., Heizmann, S., Bedner, K. Giss, D., Meschke, M., Zumb\"{u}hl, DM., Zimmerman, JD. \&. Gossard, AC. 2014 Silver-epoxy microwave filters and thermalizers for millikelvin experiments. \textit{Appl. Phys. Lett.} \textbf{104}, 211106.

\bibitem{Prance:2015} Prance, DR., Bradley, DI., George, RE., Haley, RP., Pashkin, YuA., Sarsby, M., Penttil\"{a}, J., Roschier, L., Gunnarsson, D., Heikkinen, H. \& Prunnila, M. 2015 Nanoelectronic thermometers optimised for sub-10 millikelvin operation. \textit{arXiv:1505.07244}



\end{thebibliography}


\end{document}